# Digital Real Estate in the Metaverse: An Empirical Analysis of Retail Investor Motivations

**Abstract**: This paper investigates retail investor motivations for digital real-estate ownership in the crypto-metaverse. Utilizing a detailed financial behavior survey of metaverse landowners' intrinsic and extrinsic motivations, we apply principal component analysis to uncover four distinct motivational groups: Aesthetics and Identity, Social and Community, Speculation and Investment, and Innovation and Technology. Our findings reveal that age, education, investment knowledge, risk-taking, and impulsivity significantly influence investor group membership. This research provides valuable insights to investors and developers, underscoring the potential of a platform to attract retail investors with speculative intentions, engagement longevity, and passive/active trading characteristics, contingent on unique crypto-metaverse characteristics.

## 1   Introduction

The emergence of the metaverse has catalyzed the rapid acquisition and development of digital real estate (hereafter referred to as "land"), with ownership recorded on the blockchain via non-fungible tokens (NFTs).[1] The digital land facilitates engagement and capitalization for investors and businesses within virtual worlds and gaming platforms (Davis et al., 2009; Lee et al., 2021). It encompasses diverse applications such as the commercialization of goods and services (e.g., virtual storefronts for Nike and Gucci on Roblox); the provision of immersive experiences (e.g., virtual tours of the Metropolitan Museum of Art in the virtual world of Roblox); the orchestration of events (e.g., virtual concerts featuring Snoop Dogg on the Sandbox metaverse, and virtual fashion shows by Dolce & Gabbana on Decentraland); and promotional endeavors (e.g., virtual Coca-Cola vending machines on Roblox). JP Morgan (2022) notably embarked on metaverse ventures, anticipating a market potential surpassing USD 1 trillion.

In recent years, the scholarly literature has examined the role of NFTs within the cryptocurrency ecosystem; however, the more limited subset of studies delving into metaverse land ownership via NFTs predominantly centers on determinants impacting land valuations and their interconnections with cryptocurrency markets.[2] Specifically, Dowling (2021) notes a consistent escalation in Decentraland land prices, despite pricing inefficiencies. Conversely, Goldberg et al. (2021) and Yencha (2023) link Decentraland land price premiums to proximity to metaverse landmarks, streets, and memorable address names, paralleling pricing dynamics in tangible real estate markets. A related investigation posits that Decentraland land sale advertisements emphasizing proximity to landmarks garner greater interest from potential buyers (Guidi and Michienzi, 2022). In the Sandbox metaverse, investors are more inclined to apply premiums to land sales conducted in the native cryptocurrency SAND (Nakavachara and Saengchote, 2022). The availability of new Sandbox land further bolsters the network value of existing land (Saengchote et al., 2023). Within the broader NFT market, Ante (2022) identifies interdependent causal relationships among collectible, art, and metaverse NFT markets. At the regulatory level, Goanta (2020) contends that the European Digital Content Directive may help safeguard consumer interests within the domain of digital real estate.

---

[1] The terminology used to refer to a discrete unit of digital real estate differs across metaverse platforms, encompassing terms such as "land", "parcel", "estate", and "plot", amongst others. We refer to these terms as "land" throughout this paper to facilitate brevity, clarity, and consistency. A non-fungible token (NFT) on a blockchain represents each land unit's ownership, enabling transactions and preventing replication.

[2] For a comprehensive overview of NFTs in the context of financial economics, we refer the reader to the NFT review agenda of Baals et al. (2022).

This paper represents a pioneering effort to examine the motivations driving retail investors' decisions to purchase metaverse land, advancing upon previous studies by shifting the focus from pricing dynamics to underlying factors. Decentraland and Sandbox, the largest metaverse platforms by market capitalization and users, had an average floor price of $1,203 and $1,675, respectively, for a unit of digital land.[3] Given the significant investment required, it is crucial to understand the motivations behind these purchasing behaviors. This study offers valuable insights into the market behavior of retail investors in the metaverse, identifies the underlying factors that drive investment decisions in these unique digital assets, and uncovers the motivations behind retail investor's purchasing behaviors.

This paper represents the first analysis of the motivations driving retail investor demand for metaverse land. Firstly, we conducted a survey among retail investors to determine the intrinsic and extrinsic motivations underlying their ownership of metaverse land. Secondly, utilizing principal component analysis, we identified four distinct investor groups based on their underlying motivations. Thirdly, we examine the background of each investor group and analyze the demographics that are more likely to belong to each motivation group.

---

[3] Calculated using floor price data from (CoinGecko, 2023) between December 15, 2022 and January 19, 2023. The floor price is calculated on the basis of the lowest-priced NFT within an NFT project, e.g., the "buy now" price on NFT marketplaces such as OpenSea.

The remainder of the paper is structured as follows: Section 2 delves into the conceptual background that underpins the dataset. Section 3 comprehensively describes the dataset and empirical methodology employed. Section 4 presents the results and empirical findings. Finally, Section 5 offers a conclusion, summarizing the key insights and implications of the study.

## 2    Conceptual background

Motivation plays a crucial role in shaping human behavior, directing individuals toward their goals, and influencing their interactions with their environment. Self-Determination Theory (SDT), a comprehensive macro theory of human motivation, asserts that motivation's quality and strength depend on satisfying one's basic psychological needs (Deci and Ryan, 1985).[4] SDT distinguishes between "intrinsic motivation", which stems from inherent satisfaction and pleasure, and "extrinsic motivation", which centers on achieving separable outcomes such as rewards or avoiding punishment (Ryan and Deci, 2000). In the context of retail investor ownership of metaverse land, intrinsic factors such as (i) aesthetics, (ii) utility, and (iii) identity, along with extrinsic factors including (iv) social rewards, (v) disruption, (vi) technological innovation, (vii) business models, (viii) community rewards, (ix) short-term sales, (x) long-term sales, (xi) saving, and (xii) staking opportunities, influence users' engagement with digital assets.

Intrinsic factors may attract retail investors to metaverse land ownership for several reasons. First, investors may value the (i) aesthetics of the land or project, such as visual or artistic aspects (Kong and Lin, 2021; Wang et al., 2022). Second, the (ii) identity motive emphasizes

---

the importance of self-expression and personal identity in the digital space, with users employing land and thus NFTs as a means to express their online identity through profile pictures, avatars, and other digital representations (Far et al., 2022; Turkle, 1999; Wood et al., 2014). Third, the inherent (iii) utility and satisfaction derived from interacting with the metaverse platform may draw retail investors to metaverse land ownership (Belk et al., 2022; Zhang, 2023).

Extrinsic factors can also entice retail investors to purchase metaverse land. (iv) Social rewards, such as promoting sustainability or philanthropy, and external rewards linked to contributing to the greater good can attract investors to metaverse land ownership (Casale-Brunet et al., 2022; Chandra, 2022). Intangible external rewards associated with (v) disruption – the desire to challenge traditional norms and systems through decentralization and anonymity – can also motivate land acquisition. The attraction of being an investor in (vi) technological innovations or innovative (vii) business models may yield personal external rewards (Baytaş et al., 2022; Chalmers et al., 2022). The sense of belonging to a (viii) community of like-minded individuals can also motivate retail investors, offering community relationships as an external reward (Guo and Barnes, 2007; Ridings and Gefen, 2006). The potential of extrinsic financial rewards, such as the ability to sell the land in the (ix) short-term or (x) long-term or store value through (xi) saving, can also serve as an incentive to acquire metaverse land, particularly for higher-risk speculative investments (Fisch et al., 2021; Wilson et al., 2022). Finally, the ability to stake assets can motivate retail investors by offering external rewards such as participation in governance and decision-making in metaverse projects.[5] In metaverse projects governed by decentralized autonomous organizations (DAOs), investors may receive equity, voting rights, or financial rewards through staking participation (van Haaften-Schick and Whitaker, 2022; Whitaker and Kraeussl, 2018).

---

[5] Staking in the context of crypto assets refers to the process of locking up a certain amount of a network's native tokens to, e.g., secure the network, validate transactions, and maintain network stability. In return, users who stake their tokens may earn rewards, such as additional tokens, for their contribution to the network's operations. For example, land in The Sandbox can be staked to obtain rewards in the project's native token SAND (The Sandbox, 2022).

# 3 Data background

The comprehensive dataset originates from an extensive online survey designed by the authors and distributed by coingecko.com, a leading cryptocurrency data service provider, via Twitter, Facebook, and a daily newsletter between December 15, 2022, and January 19, 2023.[6] We informed the participants – retail investors – that the primary objective was to anonymously gather insights into NFT, stablecoin, and cryptocurrency usage. Due to the survey's anonymous nature, participants were not offered any remuneration for their contributions. Figure 1 illustrates that the survey received 2,892 clicks, with 438 participants completing it. [7] Out of these retail investors, 343 disclosed current or past ownership of NFTs, and a subset of 164 owned or had previously owned metaverse land.

**--- Insert Figure 1 ---**

Our survey employed a blend of established items from existing literature and newly created items. We subjected these items to a rigorous pre-testing phase, including cognitive interviewing and pilot testing, which helped refine and clarify them. We further bolstered the reliability of the new items through robustness checks, such as internal consistency assessment and factor analysis, validating their dependability. These comprehensive steps enhanced the methodological rigor of our survey and the trustworthiness of our findings. To determine land ownership, we queried survey participants about ownership of land in prominent blockchain metaverses, including *The Sandbox*, *Decentraland*, and *Otherside*. We used this information to generate a dummy variable for land ownership. As such, individuals who might own land in other blockchain metaverses, not mentioned in our survey, are not reflected in this variable.

---

[6] The *Alternative.me cryptocurrency greed and fear index* and *Sentix Bitcoin Sentiment Index* did not substantially change throughout this period (Alternative.me, 2023; Sentix, 2023).

[7] CoinGecko users are predominantly male, younger, and with a concentration in the United States (Similar Web, 2023). Our survey demographic closely aligns with these findings. Moreover, this profile corresponds with wider demographic traits reported in the cryptocurrency market literature, such as a skewness towards male participants, younger age groups, and higher education levels (Ante et al., 2022; Auer and Tercero-Lucas, 2022; Hackethal et al., 2022; Oksanen et al., 2022; Steinmetz et al., 2021; Weber et al., 2023). These corroborating demographics underscore the representativeness and relevance of our survey sample.

Table 1 delineates the descriptive statistics for the intrinsic (i-iii) and extrinsic (iv-xii) factors, with the underlying motivations outlined in Section 2. Further, it displays results of a two-sample t-test comparing land owners with non-land owners. Following Fisch et al's. (2021) investigation of initial coin offerings, participants appraised the salience of these factors employing a Likert scale that spanned from "not important" (1) through "neutral" (3) to "very important" (5).[8] Furthermore, the survey collected data to control for demographic attributes, including gender, age, income, education, and geography, as well as investor psychological aspects such as risk propensity, impulsivity, and investment acumen. The survey also collected data to control for NFT cognizance, covering variables such as the number of NFTs owned, apprehension of fraud, regulatory necessity, project familiarity, NFT comprehension, and the timing of initial NFT acquisition. These control variables draw upon research addressing gender disparities (Bannier et al., 2019), the spectrum of risk aversion (Borri et al., 2022), and the prevalence of fraudulent activities within the cryptocurrency domain (Kshetri, 2022).

**--- Insert Table 1 ---**

Metaverse land owners, when compared with non-land NFT owners, are typically more seasoned participants in the NFT ecosystem, holding a larger portfolio of NFTs, and are predominantly from Asia. These individuals place a higher value on aesthetics, social attributes, the potential for short-term sales, and the saving capacity of NFTs. We employed a logistic regression model to delve deeper into these observations and predict land ownership among NFT owners. This analysis revealed significant negative associations with age and concerns about fraud. In contrast, we observed positive associations with the volume of NFTs owned and the year of the first NFT purchase (Appendix Table A.1).

---

[8] Fisch et al. (2021) rely on the nine motives *Utility*, *Social*, *Disruption*, *Technology*, *Business model*, *Sale (short-term)*, *Sale (long-term)*, *Equity stake* and *Financial gains* for their analysis of ICO investors, which in turn were inspired by previous studies on crowdfunding (Pierrakis, 2019; Ryu and Kim, 2016). Our extended list incorporates the same, somewhat adjusted, variables and further includes the NFT-specific motives *Aesthetics*, *Identity*, and *Community*.

## 4 Empirical model and results

### 4.1 Principal component factor analysis

Inspired by Fisch et al. (2021), we adopt an econometric methodology employed in related studies with analogous objectives (e.g., Ante et al., 2023; Ryu and Kim, 2016). Specifically, we implement exploratory factor analysis (EFA) to identify the latent structure underlying the data and categorize retail investor motivations for metaverse land ownership, as detailed in Table 1 and Sections 2–3. This approach relies on principal component analysis, employing varimax rotation. We adopt a threshold of 0.4 for assigning factor loadings to a factor in line with related studies utilizing this methodology (McCain, 1990; Peterson, 2000). Table 2 displays the results of the factor analysis. The model reveals four distinct factors, each possessing eigenvalues exceeding one. These factors account for 78.3% of the variance in our dataset.[9]

**--- Insert Table 2 ---**

Factor 1 (*Aesthetics & Identity*) comprises the variables (i) aesthetics (0.84), (iii) identity (0.69), and (vii) business model (0.69), accounting for 26.1% of the total variance. Individuals who prioritize these factors may perceive metaverse real estate as an extension of their digital personae, reflecting their aesthetic preferences, cultural proclivities, and identity (Turkle, 1999; Wang et al., 2022). The manifestation of online identity and aesthetic perception can occur through sharing digital assets on social media platforms, developing virtual properties, and integrating imagery into the cartographic attributes of virtual domains (Wood et al., 2014).

---

[9] The statistical analysis indicates that the selected research methodology and dataset are appropriate for our research project, as indicated by Bartlett's test for sphericity of (p < 0.000) and the sampling adequacy of the Kaiser-Meyer-Olkin (KMO) measure (=0.915). To control for potential early and late respondent bias, we analyzed the differences between the split sample. The independent sample t-test discovered no statistically significant deviations in the mean values of our constructs.

Factor 2 (*Social & Community*) encompasses (ii) utility (0.68), (iv) social (0.74), (viii) community (0.58), (xi) saving (0.74), and (xii) stake (0.64), accounting for 24.5% of the variance. This highlights the importance of social and community engagement within the metaverse, where retail investors may passively purchase digital real estate through saving or staking to interact with the community. This aligns with Guo and Barnes (2007), who contend that social influence is crucial in acquiring or creating possessions to enhance personal competence and social standing. In accordance with Ridings and Gefen (2006), individuals may acquire digital real estate to partake in community activities, exchange knowledge, network with peers, and establish social connections. Furthermore, it can be inferred that individuals invest in digital real estate with the intention of actively engaging in community events, knowledge-sharing, peer networking, and nurturing social connections.[10]

Factor 3 (*Speculation & Investment*) consists of positive loadings of (ix) short-term sales (0.87) and (x) long-term sales (0.81), accounting for 15.7% of the variance. Individuals within this category primarily pursue monetary gains in both short and long-term durations by trading metaverse real estate. The identification of a financially motivated group is consistent with Fisch et al. (2021); however, our analysis demonstrates positive factor loadings for both sales variables across all groups, which contrasts with ideologically motivated investors who displayed negative factor loadings for sales.

---

[10] The seemingly counter-intuitive association in our EFA between "Saving" and the "Social and Community" factor, rather than "Speculation and Investment," merits clarification. In the context of NFT land ownership, "Saving" indicates participants' inclination for long-term holding, often accompanying heightened community engagement due to a vested interest in the sustained growth of their respective metaverses. In contrast, "Speculation and Investment" pertains to those pursuing short-term, speculative profits, demonstrating a distinct divergence in investment horizons and community engagement from the "Saving" construct.

Factor 4 (*Innovation & Technology*) encompasses (v) disruption (0.82) and (vi) technology (0.69), accounting for 12% of the variance. Motivations for purchasing metaverse land within this group of exclusively extrinsic motivations may stem from technological innovations and the disruptive potential of digital realms. This group likely comprises early adopters who utilize the metaverse as an experimental environment to develop novel user experiences and investigate the underlying technology.

Table 3 presents the correlations between the four factors. The most substantial significant positive correlation (0.55) is observed between the first two factors, indicating that metaverse land owners may prioritize their real estate as a vehicle for self-expression, both visually (factor 1) and socially (factor 2).

**--- Insert Table 3 ---**

*4.2   Examining group membership dynamics*

In order to add an additional layer to our investigation, we implement an analysis that employs factor scores as dependent variables, with the objective of scrutinizing the influence of various independent variables delineated in Table 1 on the group membership dynamics of metaverse land owners. The regression models presented in Table 4 utilize the respective factor scores of the four groups as dependent variables, consequently revealing the associations between independent variables and individuals' alignments within each of the four discrete groups.[11]

---

[11] Correlation analysis and variance inflation factors (VIFs) showed that factor scores are uncorrelated and all presented variables are suitable for the regression analysis. The largest significant correlations between all variables in the regression models are r = 0.51 for NFT Tech and Project Knowledge and r = 0.41 for risk-taking and impulsivity.

Our findings demonstrate that older retail investors exhibit a significantly greater proclivity toward factor 2 (*Social & Community*) and factor 4 (*Innovation & Technology*). Moreover, investors with advanced educational backgrounds are less predisposed to align with factor 2, suggesting that such individuals who possess metaverse land may ascribe diminished importance to the social and communal dimensions inherent to metaverse land ownership. Retail investors displaying heightened risk-taking propensity and impulsivity are significantly more inclined to associate with factor 3 (*Speculation & Investment*), while concurrently exhibiting a reduced likelihood of belonging to factor 2. Retail investors possessing superior investment knowledge demonstrate a significant predilection for factor 2, although intriguingly, they do not show a similar disposition toward factors 3 and 4. Lastly, investors who are well-informed about NFTs are significantly more likely to be members of factor 1 (*Aesthetics & Identity*), implying that individuals who own metaverse land and actively pursue NFT-related knowledge may accord greater weight to the visual and distinctive characteristics emblematic of metaverse land ownership.

## 5   Conclusion

This paper contributes a pioneering examination of the multifaceted motivations underpinning digital real estate ownership within the nascent metaverse through the lens of SDT, addressing a significant lacuna in the extant literature on crypto assets. Our investigation discerns four salient categories of motivations for virtual land ownership, namely: 1) Aesthetics & Identity, 2) Social & Community, 3) Speculation & Investment, and 4) Innovation & Technology. These emergent categories underscore the intricacy and diversity of retail investor's motives and reflect the multifarious nature of virtual environments, as well as the myriad potential applications for digital real estate. By identifying distinct motivational profiles, our findings contribute to a deeper understanding of behavioral finance, technology adoption, and digital asset valuation in virtual environments.

Our findings bear considerable practical implications. For metaverse developers and stakeholders, comprehending the motivations of metaverse owners facilitates the design of bespoke, engaging virtual experiences that accommodate specific user needs, thereby broadening the user base and enhancing satisfaction and retention. Furthermore, the insights from this analysis can inform policymakers and regulators in devising more efficacious, targeted policies for governing and managing the burgeoning digital economy. As the metaverse undergoes rapid expansion and transformation, the imperatives of understanding the driving forces behind virtual land ownership are heightened, with significant ramifications for the development of this compelling new domain.

Notwithstanding the depth of our analysis, we recognize that the metaverse's fluid nature necessitates ongoing examination. With the advent of novel technologies, platforms, and use cases, future research should strive to probe the dynamic interdependencies between investor motivations and the metaverse milieu. In pursuing such inquiries, we can further illuminate the complexities of this captivating digital frontier, thereby unlocking the metaverse's full potential and guiding its trajectory towards a more interconnected, immersive, and innovative future.

# TABLES AND FIGURES

**Figure 1. Sample selection process.**

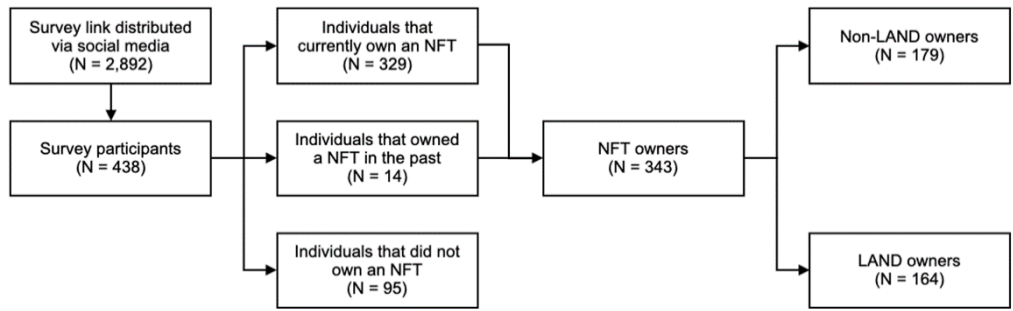

## Table 1. Descriptive statistics

| Variable | Summarized Details of Survey Question | Mean | SD | t-test | Rank |
|---|---|---|---|---|---|
| **Intrinsic Motivations** | | | | | |
| (i) Aesthetics | Acquired NFT-land for personal appreciation of artistic or aesthetic elements. Likert scale: (1) not important, to (5) very important. | 3.57 | 1.23 | -0.23* | 8 |
| (ii) Utility | Acquired NFT-land for its intended utility (e.g., governance, access, gaming). Likert scale: (1) not important, to (5) very important. | 3.65 | 1.28 | 0.15 | 3 |
| (iii) Identity | Acquired NFT-land to enhance online identity (e.g., profile pictures, avatars). Likert scale: (1) not important, to (5) very important. | 3.44 | 1.31 | -0.09 | 9 |
| **Extrinsic Motivations** | | | | | |
| (iv) Social | Acquired NFT-land for social motives (e.g., sustainability, philanthropy). Likert scale: (1) not important, to (5) very important. | 3.30 | 1.25 | -0.35** | 11 |
| (v) Disruption | Acquired NFT-land to foster disruption (e.g., decentralization, anonymity). Likert scale: (1) not important, to (5) very important. | 3.42 | 1.21 | -0.18 | 10 |
| (vi) Technology | Acquired NFT-land to explore new technology. Likert scale: (1) not important, to (5) very important. | 3.64 | 1.22 | -0.20 | 4 |
| (vii) Business model | Acquired NFT-land to engage with innovative business models or ideas. Likert scale: (1) not important, to (5) very important. | 3.62 | 1.30 | -0.18 | 5 |
| (viii) Community | Acquired NFT-land to become part of a community or network. Likert scale: (1) not important, to (5) very important. | 3.60 | 1.91 | -0.19 | 7 |
| (ix) Sale (short term) | Acquired NFT-land with the intention to sell at a higher price shortly after purchase. Likert scale: (1) not important, to (5) very important. | 3.60 | 1.30 | -0.26* | 6 |
| (x) Sale (long term) | Acquired NFT-land with the intention to sell at a higher price in the long term. Likert scale: (1) not important, to (5) very important. | 3.89 | 1.23 | -0.16 | 1 |
| (xi) Saving | Acquired NFT-land for saving purposes (e.g., inflation hedging, capital preservation). Likert scale: (1) not important, to (5) very important. | 3.22 | 1.33 | -0.34** | 12 |
| (xii) Staking | Acquired NFT-land to participate in staking or DAO (e.g., equity, voting rights, rewards). Likert scale: (1) not important, to (5) very important. | 3.70 | 1.22 | -0.19 | 2 |
| **Demographics** | | | | | |
| Gender | Gender. Dummy variable: (1) male, (0) female. | 0.87 | - | 0.04 | - |
| Age | Age. Categorical variable: (1) 18 to 24; (2) 25 to 34; (3) 35 to 44; (4) 45 to 54; (5) 55 to 64; (6) 65 and over. | 2.54 | 1.05 | 0.35*** | - |
| Net income | Average monthly net income. Categorical variable: (1) less than $500; (2) $500 to $999; (3) $1,000 to $1,499; (4) $1,500 to $1,999; (5) $2,000 to $2,999; (6) $3,000 to $4,999; (7) more than $5,000. | 2.88 | 2.45 | 0.50* | - |
| Education | Highest education. Categorical variable: (1) no schooling; (2) high school, trade, technical, vocational training, apprenticeship; (3) Bachelor's degree; (4) Master's degree; (5) Doctorate degree. | 2.54 | 1.22 | 0.02 | - |
| Continent: North America | Located in North America. Dummy variable: (1) yes, (0) no. | 0.13 | - | 0.14*** | - |
| Continent: Asia | Located in Asia. Dummy variable: (1) yes, (0) no. | 0.32 | - | -0.12*** | - |
| Continent: Australia | Located in Australia. Dummy variable: (1) yes, (0) no. | 0.03 | - | 0.01 | - |
| Continent: South America | Located in South America. Dummy variable: (1) yes, (0) no. | 0.05 | - | -0.02 | - |
| Continent: Africa | Located in Africa. Dummy variable: (1) yes, (0) no. | 0.21 | - | -0.06 | - |
| Continent: Europe | Located in Europe. Dummy variable: (1) yes, (0) no. | 0.26 | - | 0.06 | - |
| **Investor Psychology Traits** | | | | | |
| Risk-Appetite | Risk-taking willingness. Likert scale: (1) not willing, to (10) very willing. | 7.65 | 2.30 | -0.13 | - |
| Impulsivity | Impulsivity level. Likert scale: (1) not impulsive, to (10) very impulsive. | 5.89 | 2.50 | -0.28 | - |
| Investment Knowledge | Non-crypto investment knowledge. Likert scale: (1) no knowledge, to (10) very knowledgeable. | 6.73 | 2.34 | -0.17 | - |

**NFT Cognizance**

| | | | | | |
|---|---|---|---|---|---|
| Number of NFTs owned | Number of NFTs owned. Categorical variable: (1) 0; (2) 1; (3) 2-5; (4) 6-10; (5) 11-20; (6) 21-50; (7) 51 or more. | 4.26 | 1.81 | -0.64*** | - |
| Fear of NFT fraud | Concern about NFT fraud. Likert scale: (1) strongly disagree, to (5) strongly agree. | 3.45 | 1.20 | 0.14 | - |
| NFT regulation desire | Need for stricter NFT regulation. Likert scale: (1) strongly disagree, to (5) strongly agree. | 3.37 | 1.31 | -0.09 | - |
| NFT project knowledge | Awareness of NFT project before investment. Likert scale: (1) strongly disagree, to (5) strongly agree. | 3.84 | 1.08 | 0.01 | - |
| NFT tech knowledge | Familiarity with NFT technology before investment. Likert scale: (1) strongly disagree, to (5) strongly agree. | 3.80 | 1.16 | 0.03 | - |
| NFT first purchase | Year of first NFT purchase. Categorical variable: (1) 2017; (2) 2018; (3) 2019; (4) 2020; (5) 2021; (6) 2022. | 2.48 | 1.20 | -0.33** | - |

*Table 1 reports descriptive statistics from an online survey designed by the authors and distributed by coingecko.com, a leading cryptocurrency data service provider, via Twitter, Facebook, and a daily newsletter between December 15, 2022, and January 19, 2023. The column t-test indicates the results of a t-test, i.e., the difference in comparison to NFT owners that do or did not own any metaverse land. Statistical significance at the 1%, 5% and 10% levels are denoted by ***, ** and *, respectively.*

**Table 2. Profiling metaverse land ownership using principal components analysis**

| | Factor 1:<br>Aesthetics and Identity | Factor 2:<br>Social and Community | Factor 3:<br>Speculation and Investment | Factor 4:<br>Innovation and Technology |
|---|---|---|---|---|
| Aesthetics | **0.84** | 0.12 | 0.21 | 0.24 |
| Identity | **0.76** | 0.39 | 0.12 | 0.05 |
| Business model | **0.69** | 0.29 | 0.24 | 0.35 |
| Social | 0.28 | **0.74** | 0.10 | 0.41 |
| Saving | 0.14 | **0.74** | 0.31 | 0.24 |
| Utility | 0.55 | **0.68** | 0.10 | 0.07 |
| Stake | 0.41 | **0.64** | 0.40 | -0.01 |
| Community | 0.52 | **0.58** | 0.06 | 0.37 |
| Sale (short-term) | 0.18 | 0.12 | **0.87** | 0.01 |
| Sale (long-term) | 0.12 | 0.23 | **0.81** | 0.25 |
| Disruption | 0.31 | 0.27 | 0.19 | **0.82** |
| Technology | 0.42 | 0.49 | 0.13 | **0.69** |
| Variance explained | 26.1% | 24.5% | 15.7% | 12.0% |
| Cronbach's α | 0.77 | 0.83 | 0.77 | 0.71 |

*Table 2 presents the findings of exploratory factor analysis focused on metaverse real estate owners – retail investors. Factor loadings allocated to the respective factors are accentuated in bold. The factor analysis is executed employing principal component analysis and Varimax rotation with Kaiser normalization, specifically for metaverse land owners. N = 164. Kaiser-Meyer-Olkin (KMO) measure: 0.915; Bartlett's test of sphericity: p < .000. Harman's one-factor test for common method bias: 0.38.*

**Table 3. Correlations between metaverse landowner factors**

| Groups / factors | (1) | (2) | (3) | (4) |
|---|---|---|---|---|
| (1) Aesthetics and Identity | 1.00 | | | |
| (2) Social and Community | 0.55 | 1.00 | | |
| (3) Speculation and Investment | 0.35 | 0.32 | 1.00 | |
| (4) Innovation and Technology | 0.39 | 0.36 | 0.23 | 1.00 |

*Table 3 displays the discerned correlations between the four factors extracted from an exploratory factor analysis employing oblimin rotation for a sample comprising metaverse real estate owners – retail investors. All coefficients exhibit significant correlations at the 1% level. The factor analysis is conducted using principal component analysis and Oblimin rotation, specifically for metaverse land owners. N = 164. Kaiser-Meyer-Olkin (KMO) measure: 0.915; Bartlett's test of sphericity: p < .001.*

**Table 4. Regression models predicting metaverse land owner profiles**

| | Factor 1: Aesthetics and Identity | Factor 2: Social and Community | Factor 3: Speculation and Investment | Factor 4: Innovation and Technology |
|---|---|---|---|---|
| **Demographics** | | | | |
| Gender (male) | -0.15 (0.37) | 0.23 (0.24) | 0.19 (0.34) | 0.05 (0.33) |
| Age | 0.05 (0.08) | 0.16 (0.07)** | -0.03 (0.07) | 0.14 (0.07)** |
| Net income | -0.02 (0.04) | -0.06 (0.04) | 0.02 (0.04) | 0.07 (0.04) |
| Education | 0.12 (0.08) | -0.17 (0.07)** | -0.01 (0.07) | -0.01 (0.07) |
| **Investor Psychology Traits** | | | | |
| Risk-appetite | -0.01 (0.04) | 0.07 (0.04) | 0.08 (0.04)** | -0.06 (0.04) |
| Impulsivity | 0.03 (0.04) | -0.06 (0.03)* | 0.08 (0.03)** | -0.01 (0.04) |
| Investment knowledge | -0.03 (0.05) | 0.08 (0.04)** | -0.05 (0.05) | 0.05 (0.04) |
| **NFT Cognizance** | | | | |
| Number of NFTs owned | -0.05 (0.05) | 0.04 (0.05) | -0.01 (0.05) | 0.04 (0.04) |
| Fear of NFT fraud | 0.01 (0.07) | -0.07 (0.07) | 0.07 (0.07) | 0.02 (0.09) |
| NFT regulation desire | -0.03 (0.07) | 0.08 (0.08) | 0.08 (0.07) | -0.09 (0.08) |
| NFT project knowledge | 0.23 (0.14)* | 0.03 (0.10) | 0.10 (0.11) | 0.16 (0.12) |
| NFT tech knowledge | 0.11 (0.14) | 0.11 (0.10) | 0.06 (0.11) | 0.07 (0.10) |
| NFT first purchase | 0.02 (0.06) | 0.02 (0.06) | 0.06 (0.07) | -0.04 (0.06) |
| F-statistic | 0.004 | 0.000 | 0.000 | 0.016 |
| $R^2$ (Adj. $R^2$) | 0.14 (0.04) | 0.26 (0.19) | 0.26 (0.18) | 0.17 (0.05) |

*Table 4 presents the outcomes of four ordinary least squares regression models with heteroskedasticity-robust standard errors and a constant term (not reported) to predict factor scores for the four retail investor groups. N = 164. Statistical significance at the 5% and 10% levels are denoted by ** and *, respectively.*

**Table A.1. Regression model predicting metaverse land ownership**

| | Odds ratio | SE | z | 95% CI | |
|---|---|---|---|---|---|
| | | | | Lower | Upper |
| **Demographics** | | | | | |
| Gender (male) | 0.523 | 0.214 | -1.59 | 0.235 | 1.165 |
| Age | 0.808 | 0.087 | -1.98** | 0.655 | 0.998 |
| Net income | 0.964 | 0.055 | -0.64 | 0.862 | 1.078 |
| Education | 0.973 | 0.105 | -0.25 | 0.787 | 1.202 |
| **Investor Psychology Traits** | | | | | |
| Risk-appetite | 0.989 | 0.065 | -0.17 | 0.868 | 0.870 |
| Impulsivity | 1.026 | 0.057 | 0.47 | 0.640 | 0.921 |
| Investment knowledge | 1.057 | 0.059 | 0.99 | 0.322 | 0.947 |
| **NFT Cognizance** | | | | | |
| Number of NFTs owned | 1.192 | 0.082 | 2.55** | 0.011 | 1.041 |
| Fear of NFT fraud | 0.766 | 0.093 | -2.19** | 0.029 | 0.603 |
| NFT regulation desire | 1.158 | 0.129 | 1.32 | 0.186 | 0.932 |
| NFT project knowledge | 0.948 | 0.139 | -0.37 | 0.714 | 0.710 |
| NFT tech knowledge | 1.024 | 0.142 | 0.17 | 0.866 | 0.780 |
| NFT first purchase | 1.230 | 0.127 | 2.01** | 0.045 | 1.005 |

*Table A.1 presents the outcomes of a logistic regression model and a constant term (not reported) to predict metaverse land ownership across NFT owners. The model includes a constant and controls for continents where NFT users are located (not reported). N = 343. Pseudo $R^2$ = 0.10; $Chi^2$ = 0.0003. Statistical significance at the 5% levels are denoted by **.*